%% file: ms.tex
\documentclass{sig-alternate}

\usepackage{color}

\newcommand\figref[1]{Figure~\ref{fig:#1}}
\newcommand\secref[1]{Section~\ref{sec:#1}}

\newcommand{\mypara}[1]{\medskip\noindent\textbf{#1}}
\newcommand{\takeaway}[1]{\medskip\noindent\textbf{\em #1}}

\begin{document}
%
% --- Author Metadata here ---
%\conferenceinfo{DAC}{'15 San Francisco, California USA}
%\CopyrightYear{2007} % Allows default copyright year (20XX) to be over-ridden - IF NEED BE.
%\crdata{0-12345-67-8/90/01}  % Allows default copyright data (0-89791-88-6/97/05) to be over-ridden - IF NEED BE.
% --- End of Author Metadata ---

\title{Address Translation Design Tradeoffs \\ for Heterogeneous Systems}
%% \subtitle{[Extended Abstract]
%% \titlenote{A full version of this paper is available as
%% \textit{Author's Guide to Preparing ACM SIG Proceedings Using
%% \LaTeX$2_\epsilon$\ and BibTeX} at
%% \texttt{www.acm.org/eaddress.htm}}}


\author{Yunsung Kim\titlenote{yunsung.kim@columbia.edu} \and Guilherme Cox\titlenote{cox@computer.org} \and Martha A. Kim\titlenote{martha@cs.columbia.edu} \and Abhishek Bhattacharjee\titlenote{abhib@cs.rutgers.edu}}

\maketitle
\begin{abstract}
\input{abstract}
\end{abstract}

%\category{B.3.2}{Hardware}{Memory Structures}{Design Styles}[Virtual memory]
%\category{C.1.3}{Computer Systems Organization}{Processor Architectuers}{Other Architecture Styles}[Heterogeneous (hybrid) systems]

\terms{Design, Measurement, Performance}

%\keywords{\fixme{ACM proceedings, \LaTeX, text tagging}}

%% 
%% INTRO
%% TAX
%% DESIGN SPACE EXPLORATION
%%     methodology
%%     experiment - motivate the rooted MMU for accelerators (can't go back to CPU; comm kills you in perf and energy; techniques in 4.2 4.3 don't mitigate it)
%%     experiment - MMU for accels design space
%%     experiment - system-wide MMU design (area v. energy optimality)
%% RECOMMENDATIONS/FRAMEWORK (coherence? OS issues?)
%% CONCL
%% 

\input{intro}

\input{tax}

\input{dse}

\input{recommend}

%\input{concl}

%% %ACKNOWLEDGMENTS are optional
%% \section{Acknowledgments}
%% This section is optional; it is a location for you
%% to acknowledge grants, funding, editing assistance and
%% what have you.  In the present case, for example, the
%% authors would like to thank Gerald Murray of ACM for
%% his help in codifying this \textit{Author's Guide}
%% and the \textbf{.cls} and \textbf{.tex} files that it describes.

\bibliographystyle{abbrv}
\bibliography{heteroTLB}  % sigproc.bib is the name of the Bibliography in this case
\end{document}

%% file: abstract.tex
This paper presents a broad, pathfinding design space exploration of
memory management units (MMUs) for heterogeneous systems. We consider
a variety of designs, ranging from accelerators tightly coupled with
CPUs (and using their MMUs) to fully independent accelerators that
have their own MMUs.  We find that regardless of the CPU-accelerator
communication, accelerators should not rely on the CPU MMU for any
aspect of address translation, and instead must have its own, local,
fully-fledged MMU.  That MMU, however, can and should be as
application-specific as the accelerator itself, as our data indicates
that even a 100\% hit rate in a small, standard L1 Translation
Lookaside Buffer (TLB) presents a substantial accelerator performance
overhead.  Furthermore, we isolate the benefits of individual MMU
components (e.g., TLBs versus page table walkers) and discover that
their relative performance, area, and energy are workload dependent,
with their interplay resulting in different area-optimal and
energy-optimal configurations.

%% This is the abstract I wrote, before I realized we were constrained to use
%% the exact abstract submitted at the abstract deadline.  -MK
%%
%% In the near future, heterogeneous systems will improve performance and
%% energy via many on-chip accelerators. Industry trends point to the
%% adoption of unified virtual and physical address spaces to ease the
%% programming model of such massively heterogeneous systems.  This opens
%% up the question of how best to support address translation on
%% accelerators. The design of memory management units (MMUs) for
%% heterogeneous systems -- their complexity, integration with the memory
%% subsystem, and role in CPU-accelerator communication overheads -- have
%% first-order impacts on hardware and operating system (OS) design.
%% This paper presents a broad, pathfinding design space exploration of
%% MMUs for heterogeneous systems.  Our analysis finds that MMUs should
%% be as application-specific as the compute engine they support, and
%% that the benefits and costs of individual MMU components (e.g., TLBs
%% versus page table walkers) in performance, area, and energy are
%% extremely nuanced and, again, workload-dependent.  Their interplay is
%% shown to result in different area-optimal and energy-optimal MMU
%% configurations.

%% file: intro.tex
\section{Introduction}
\label{sec:intro}

While accelerators offer much-needed gains in serial performance and
energy efficiency, their integration into heterogeneous systems is
primarily physical and their {\em logical integration} is
incomplete.  Accelerators are connected in ad-hoc ways with a
variety of processor-accelerator communication overheads.  Meanwhile,
application and system software must deal with cumbersome
synchronization interfaces, dedicated memory regions, and specialized
data formats.

This paper investigates a key aspect of logical integration: the
unification of accelerators and CPUs in a shared virtual and physical
address space.  A unified address space makes data structures and
pointers globally visible, obviates the need for expensive and
error-prone data marshaling between CPUs and accelerators, and
un-burdens CPUs from pinning accelerator data pages in fixed locations
in main memory, improving memory efficiency~\cite{vo:wivosca13}.
Recent work tries to reduce transfer times with smarter packing and
unpacking schemes ~\cite{gelado:asplos10, hechtman:ispass13,
  jablin:cgo12, jablin:pldi11}; similarly, recent CUDA releases permit
limited CPU/GPU virtual address sharing \cite{wilt:cuda}. However,
none of these approaches solve the problem as generally and as
flexibly as unified address spaces.  A unified address space, for
example, is a key component of the Heterogeneous System Architecture
\cite{kyriazis:hsa} specification promoted by the likes of AMD, Intel,
NVIDIA, Qualcomm, ARM, and Samsung.

With its heterogeneous uniform memory access (hUMA) technology
\cite{rogers:huma}, AMD's recent Berlin processor committed fully to
HSA, and the academic community has responded with memory management
unit (MMU) designs for GPUs \cite{pichai:asplos14, power:hpca14}, one
of the more mature acceleration technologies currently available.
Both studies have shown that address translation overheads in GPUs can
be brought down to the levels traditionally deemed acceptable in the
CPU world, 5-15\% of runtime.

This work goes beyond the state of the art by considering the general
design space of MMU hardware for {\em any abstract accelerator},
extracting {\em general principles of how to achieve efficient
  virtual-to-physical address translation}.  First, we taxonomize the
diverse population of accelerators, organizing it along four key
dimensions (\secref{tax}).  Second, we use those dimensions to explore
a design space of generic accelerators, extracting principles for
heterogeneous MMU design (\secref{dse}).  Finally, we close with a set
of recommendations and strategies for address translation in
heterogeneous systems (\secref{recommend}).

%%%%%%%%%%%%%%%%%%%%%%%%%%%%%%%%%%%%%%%%%%%%%%%%%%%%%%%%%%%%%%%%%%%%%%%%%%
%% 
%% ... model a range of accelerators with varying speedups, MMU designs,
%% communication costs with the CPU, and memory access
%% characteristics. We offload hotspots in a suite of benchmarks and
%% study how their load/store patterns perform on various MMU
%% configurations. In particular we study the impact of TLB capacities,
%% page table walker designs, and level of CPU-accelerator sharing of
%% MMUs, focusing primarily on performance, area, and energy, to produce
%% a set of general principles for address translation in heterogeneous
%% systems.

%% file: tax.tex
\section{Taxonomy of Accelerators}
\label{sec:tax}

Reasoning about accelerators in a general way requires careful
understanding of the space of possible accelerators.  While previous
accelerator taxonomies have focused on their programming
models~\cite{cascaval:jrd10}, we classify accelerators in a hardware-centric
way that defines the coverage and bounds of the design space
exploration that follows in~\secref{dse}.
There are four key dimensions in this feature space.

\mypara{Offload kernel size.} Accelerators run the gamut in terms of
the amount of work they accept per invocation, from large,
application-like GPGPU kernels to fine-grain operations such as vector
instructions or DCTs.  A range of engines such as
transcoders, checksums, compression blocks, and facial
recognition engines cover the middle ground.

\mypara{Kernel speedup over software.} Depending on how an
accelerator is imlemented and its degree of programmability, it will
offer different speedups on the kernel relative to a software
implementation.

\mypara{Characteristic reference pattern.} The acceleration target and
accelerator microarchitecture combine to produce a characteristic
memory reference pattern for each accelerator.  Some, such as FFT or
data analytics, will be streaming, whereas others, such as
highly-multithreaded-GPUs, will produce essentially random accesses.
The list of possible reference patterns is unbounded, but we will
cover three broad classes in our design space exploration.

\mypara{Distance from CPU.} Lastly, some accelerators are tightly
coupled with a CPU while others are more distant.  This is driven not
only by integration choices, but also by the desired degree of
autonomy from the CPU and its state.

%% file: dse.tex
\section{Design Space Exploration}
\label{sec:dse}

In this MMU design space exploration we will explicitly reason
about the impact of each of these four dimensions on MMU design, both
for the accelerator and when taking the CPU and accelerator memory
management together as a whole in a heterogeneous system MMU.

\input{methods.tex}
\input{expt1} % motivate the need for rooted MMUs for accelerators (comm kills you)
\input{expt2} % explore design space of MMUs for accelerators (only)
\input{expt3} % system-wide MMU design (area v. energy optimality)

%% file: methods.tex
\subsection{Experimental Methodology}
\label{sec:methods}

Here we lay out the key features of our design space exploration
methodology.  

\mypara{Benchmarks.}  To focus on acceleration rather than
parallelization, we use eight single-threaded benchmarks from SPEC
(applu, art, astar, bwaves, cactus, gems, mcf, and swim) and two
scientific applications (graph500 and gups).  For each application, we
simulate a 250M instruction simpoint trace.  As there are no workloads
that map to abstract accelerators, we use these general workloads,
isolating acceleration targets as described below.

\mypara{Identification of hotspots.}  To identify regions of each
benchmark that might typically be targeted by an accelerator, we
select hotspots from each trace.  For these experiments we consider an
instruction ``hot'' if it executes more than 100K times.  Our
heterogeneous MMU simulator, described below, sends uninterrupted
sequences of hot instructions longer than a {\em hotspot size
  threshold} to run on the accelerator, with all other instructions
executed by the CPU.

\begin{figure}
  \includegraphics[width=\columnwidth]{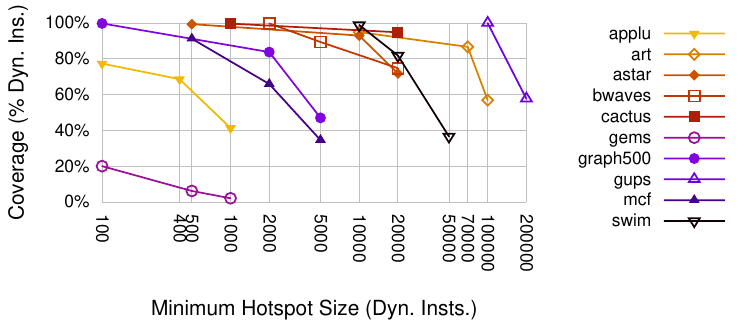}
  \caption{For each application we consider several minimum hotspot
    sizes, collectively cover three orders of magnitude in hotspot
    size.}
  \label{fig:hotspots}
\end{figure}

We consider the range of hotspot size thresholds shown in
\figref{hotspots} to capture the range of offload sizes seen in modern
accelerators.  For example, we consider three versions of applu, one
with hotspots larger than 100 dynamic instructions, another with
hotspots larger 400, and a third with hotspots larger than 1000,
resulting in acceleration of 78\%, 69\%, and 40\% respectively of the
application's dynamic instructions.  Note on the x-axis of
~\figref{hotspots} that these hotspot sizes range over three orders of
magnitude, from hundreds to tens of thousands of instructions.

\mypara{Accelerator reference patterns.}  To understand the impact an
accelerator's memory access pattern has on the MMU, the simulator can
apply a transformation to the accelerator's reference stream.  These
transformations change only the order in which the addresses are
accessed, not the addresses themselves.  Matching the breadth of
accelerators found in \secref{tax}, we use three patterns: original
program order, random order (i.e., minimal locality), and sorted
order, mimicking a streaming accelerator with high locality.

\mypara{Heterogeneous MMU simulator.}  
Unfortunately detailed CPU simulators are too slow to run workloads
large enough for virtual memory studies \cite{basu:isca13,
  bhattacharjee:micro13}, and the issue is compounded here
by our need to conduct a broad hardware design space exploration.
Thus, like in other recent work \cite{basu:isca13, bhattacharjee:micro13,
  pham:hpca14, pham:micro12}, our simulator is detailed with respect
to the dynamics of the MMU but, because the core and cache
configurations are \emph{not} part of the target design space, we
calibrate and use a first-order model.

The detailed part of the simulator models a spectrum of MMU
configurations ranging from no address translation support on the
accelerator -- requiring the accelerator to fetch all address
translations from the CPU -- to full accelerator-based translation
with a multithreaded page table walker.  As input, it takes the
instruction trace described above.

Like recent work on GPU MMUs \cite{pichai:asplos14, power:hpca14}, we
focus on dTLB accesses, which affect the system far more than iTLB
accesses.  Similarly, we assume x86-style, four-level, radix-tree page
tables, specifically focusing 4KB page sizes instead of larger (2MB
and 1GB) pages.  While large pages likely reduce TLB miss rates, they
also incur overheads (special OS algorithms for large page promotion,
increased page traffic, pinning restrictions \cite{navarro:osdi02,
  pham:micro12}) that would not be properly captured by our
experimental methods.  We also assume, like other recent MMU
work~\cite{pham:hpca14, pham:micro12, pichai:asplos14}, the absence of
page faults moving data from backing store to main memory, as most
systems have sufficient memory to eliminate page faults in the steady
state \cite{basu:isca13}.

Each TLB's access time, energy, and area are derived from CACTI
6.5~\cite{muralimanohar:micro07}, while the page table walker area
and power is characterized from RTL by Synopsys Design Compiler.  For
page table walk time, the simulator accepts a parameter derived from
real performance counter measurements.  Like past work
\cite{basu:isca13, bhattacharjee:micro13}, this is reasonable provided
that the cache configuration, which is the primary determinant of page
table walk time, does not change.  As cache configuration is not part
of our design space, it does not.

The simulator computes overall execution time as thes sum of four
terms: $Time = CPU_{non-MMU} + CPU_{MMU} + Acc_{non-MMU} + Acc_{MMU}$
The MMU times are calculated in detail via the simulated trace, while
the non-MMU terms are computed analytically: $CPU_{non-MMU} =
\frac{CPI_{App} \cdot Instrs_{CPU}}{GHz}$.  The $CPI_{App}$ term is a
simulator parameter that calibrates this model with each application's
performance counter-derived CPI when not walking the page table.  The
rationale again is that the core configuration is not part of our
design space and thus this component of performance is not subject to
significant change across our design space.  The accelerator's
computation time term is very similar to that of the CPU, but with a
speedup applied.

\mypara{Simulator calibration and validation.}  We calibrate and
validate the system model using an Intel Xeon CPU with Sandybridge
cores and a 4MB last-level cache.  Each core has a 64-entry and
512-entry L1 and L2 TLB.  On the calibration side, this is the server
on which we compute each application's CPI and average page walk time
as input parameters for the simulator.  On the validation side, we
configure the simulator to match this machine's MMU and compare how
well the simulator tracks real-system performance.  Across all 10
benchmarks, the the coefficient of determination (R squared value) of
estimated versus actual execution time of 0.955.

\mypara{Experimental baseline and normalization.}  In the experiments
that follow, much of the data will be normalized to the performance,
MMU area, and MMU energy of the Sandybridge configuration.  When
relative performance is plotted, it is the performance of the whole
application relative to unaccelerated software executing on the
Sandybridge.

%%%%%%%%%%%%%%%%%%%%%%%%%%%%%%%%%%% LOCAL BITBUCKET

%The only difference is that this equation includes a accelerator
%speedup factor, that captures the relative performance of an abstract,
%``black-box'' accelerator.

%\begin{figure}
%  \includegraphics[width=\columnwidth]{data/validate.pdf}
%  \caption{Modeled trace execution time correlates extremely strongly
%    with measured execution time on the real baseline Sandybridge
%    system.}
%  \label{fig:validate}
%\end{figure}

%% file: expt1.tex
\subsection{CPU-Managed Address Translation}
\label{sec:expt1}

We first examine the dynamics CPU-managed address translation.  Here,
accelerators invoke the CPU 
to help translate virtual addresses, so in all of these configurations
{\em the CPU is ultimately responsible for address translation}.  In
the first category, the accelerator has no MMU and probes the CPU's
MMU on every reference.  In the second, the accelerator maintains an
L1 TLB, probing the CPU only on an L1 miss.  In both cases, remote
probes of the CPU's MMU access the CPU's L2 TLB, as L1 TLBs are
tightly coupled with the CPU pipeline and would require costly
arbitration logic to coordinate CPU accesses with remote accelerator
accesses.

\begin{figure}
  \includegraphics[width=\columnwidth]{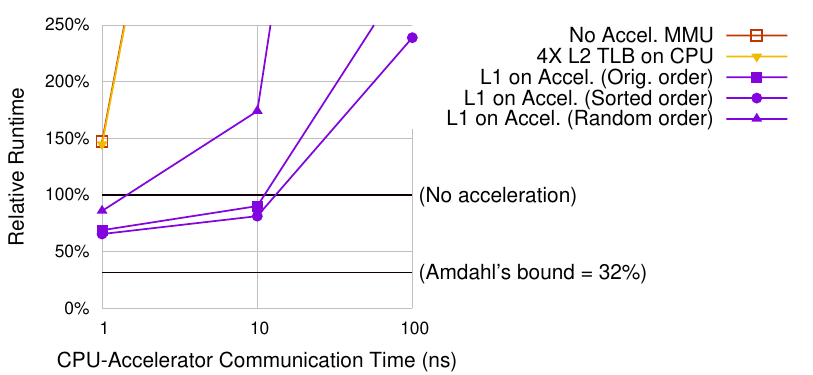}
  \caption{When an accelerator is reliant on the CPU in any way for
    address translation, the communication costs between the
    accelerator and the CPU overwhelm any benefits of acceleration.}
  \label{fig:expt1}
\end{figure}

In the following experiments, we measure the application runtime if
hotspots are offloaded to an accelerator that offers 10x speedup and
the remainder of the code is run on the CPU with the baseline MMU
configuration.  \figref{expt1} plots the results, with each datapoint
representing an average across 28 experiments -- 10 applications with
2 or 3 hotspot sizes per application.  For reference, we have
highlighted two key points: the runtime of the unaccelerated,
software-only application -- 100\% -- and the lower bound on overall
runtime given Amdahl's law -- 32\%.

\mypara{No accelerator MMU.}
We first consider scenarios where the accelerator has no MMU, and all
accelerator memory references look up the CPU L2 TLB. Intuitively,
paying a remote communication cost for every memory reference quickly
swamps the accelerator's original benefits, and this is borne out by
the data in \figref{expt1} where we see that even with an aggressively
low 1ns communication delay between the accelerator and the CPU, any acceleration benefits
are more than outweighted by the address translation activity back and
forth between accelerator and CPU.

One component of the cost of probing the CPU MMU is the
overhead of handling L2 TLB misses, so there may be system-wide
benefits to increasing CPU L2 TLB size, boosting the hit rate, and
reducing average address translation time.  The second data series in
\figref{expt1} plots the average application runtime when the CPU L2
TLB capacity is quadrupled -- from 256 to 1024 entries -- and we see
that this boost in resources has no impact on the runtimes, suggesting
that communication costs dominate any benefits of acceleration
  when the accelerator has no MMU hardware of its own.

\mypara{Accelerator with L1 TLB filter.}  Intuitively, an L1 TLB on
the accelerator would filter communication between the accelerator and
CPU. \figref{expt1} also plots the impact of a 64-entry, 4-way
set-associative L1 TLB filter -- the largest of four filters we tried.
We found that such a filter does improve performance, yet despite
achieving hit rates in excess of 96-98\% this TLB is insufficient to
eliminate communication-induced overheads.  Even under optimistic
delays of 1-10ns and reference patterns with maximal locality, address
translation eats half to three quarters of the accelerator's potential
savings.

Overall, these two analyses signal strongly that the only feasible way
to support a unified address space with realistic accelerator-CPU is
to remove CPU-accelerator communication from the picture, by
empowering an accelerator to walk the page table.  The following
experiments explore the tradeoffs in this space.

%% file: expt2.tex
\subsection{MMUs For Accelerators}
\label{sec:expt2}

\begin{figure*}
  \includegraphics{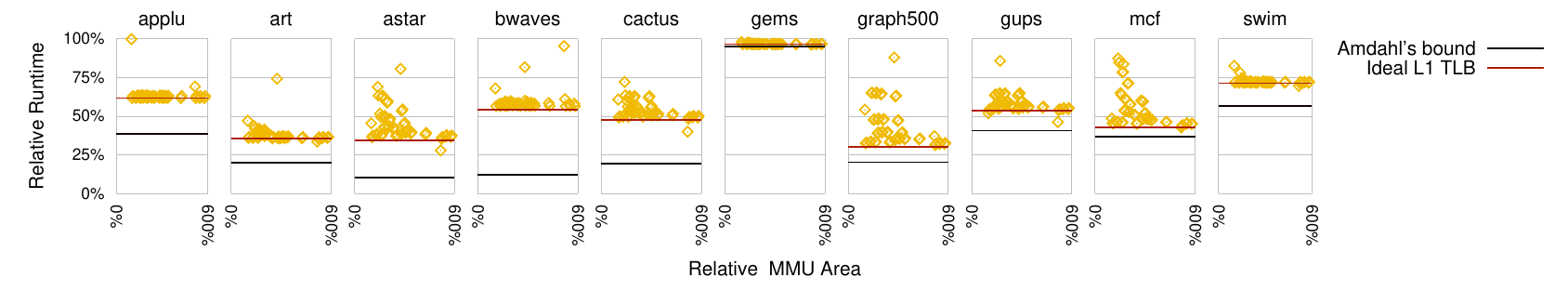}
  \caption{As area is dedicated to accelerator MMUs, some applications
    benefit significantly while others only minimally.  However, even
    the best generic MMU configuration -- an L1 TLB with 100\% hit
    rate -- leaves a significant performance gap off of Amdahl's
    acceleration ideal.}
  \label{fig:expt2}

  \includegraphics{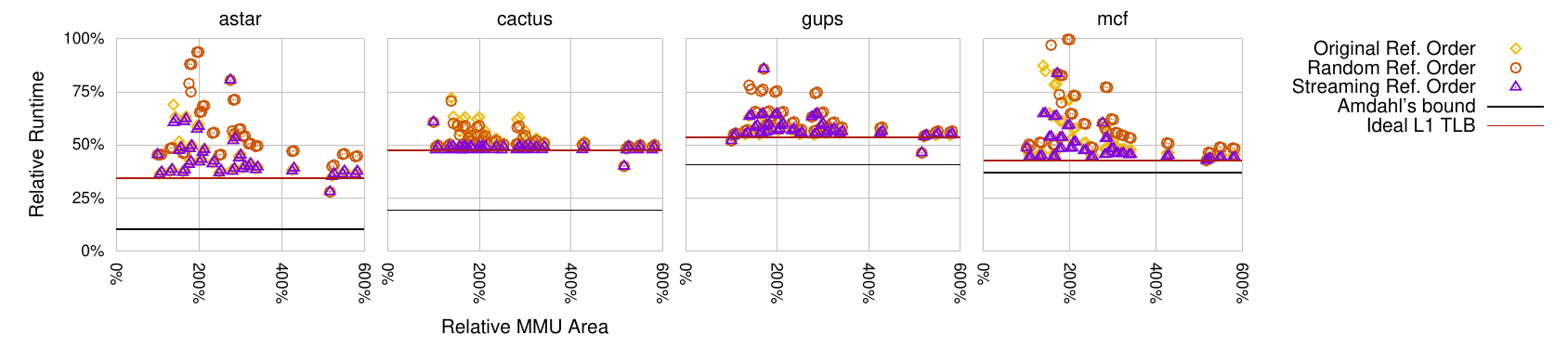}
  \caption{Benchmarks astar, cactus, gups, and mcf respond in different ways
    to different accelerator memory reference patterns.  Small
    accelerator MMUs are notably good at servicing streaming reference
    patterns, but as MMUs grow, the performance differences between
    accelerator reference patterns fade.}
  \label{fig:expt2-patts}
\end{figure*}

There are many potential MMU implementations for accelerators, and
understanding the relative importance of resources and topology (e.g.,
single-level or two-level) is critical for a resource-efficient
design.

First, the right MMU design is highly application- and reference
pattern-dependent, with a potential co-design opportunity between
candidate accelerator microarchitectures and their corresponding MMUs.
For example, is it preferable to have a smaller accelerator with
little locality and a large MMU or a larger accelerator with better
locality and a small MMU?  The answer depends on the particulars of
the application and accelerators and MMUs, suggesting that accelerator
MMUs are best viewed as specialized, application-specific circuits,
just like the accelerators they support.

Second, when given an MMU area budget, it is not at all obvious
which components -- L1 TLBs, L2 TLBs, or PTWs -- are most critical to
performance.  The answer depends on the cost-benefit of the various
components which is again application-dependent. Somewhat surprisingly, we
find that PTW organization is sometimes more critical than the other
components, a fact that is usually ignored by CPUs
\cite{bhattacharjee:micro13}, which focus on improving TLB
organization. This, in turn, suggests that there may be substantial
value in exploring the design of novel page tables, page table locking
and scalability \cite{clements:eurosys13}.

\mypara{Accelerator MMU design space.}  In this analysis, we evaluate
54 acclerator MMU configurations for each application.  While we find
that speedups and hotspots sizes influence the results slightly, the
relative benefits and trends of various MMU organizations remain
unaffected regardless of the exact accelerator speed or hotspot size.
Therefore, we will present only acceleration factors of 10x on the
middle hotspot size for each application.

The 54 MMU configurations are as follows.

\begin{itemize}
\item First there are {\em 6 page-table-walker (PTW) designs} where
  the accelerator MMU has no TLBs or other caches, just a hardware
  page table walker.  While a standard CPU PTW resolves TLB misses one
  at a time, recent work has shown the benefits of multithreaded PTWs
  that resolve multiple TLB misses together \cite{power:hpca14}, so we
  examine PTWs that resolve 1, 2, 4, 8, 16, and 32 translations at a
  time.

\item Second, we assess the benefits of placing 32-entry (2-way), and 64-entry
(4-way) L1 TLBs in tandem with PTWs of different multithreaded
factors for a total of {\em 12 L1 TLB plus PTW configurations}.

\item Finally, we incorporate a second level into the TLB hierarchy with 256, 512, and
1024-entry TLBs -- the sizes commercially available for CPUs
\cite{bhattacharjee:hpca11} -- for {\em 36 three-level MMU hierarchies}.
\end{itemize}

As additional reference points, we mark two values when presenting
these designs in \figref{expt2}, the minimum runtime as bounded by
Amdahl's law, and a slightly less optimistic scenario where every
accelerator memory reference is met with a hit in our smallest,
fastest L1 TLB (32-entry, 2-way associative).

%% HOISTED from expt3.tex
\begin{figure*}
  \begin{minipage}[t]{5in}
 \includegraphics{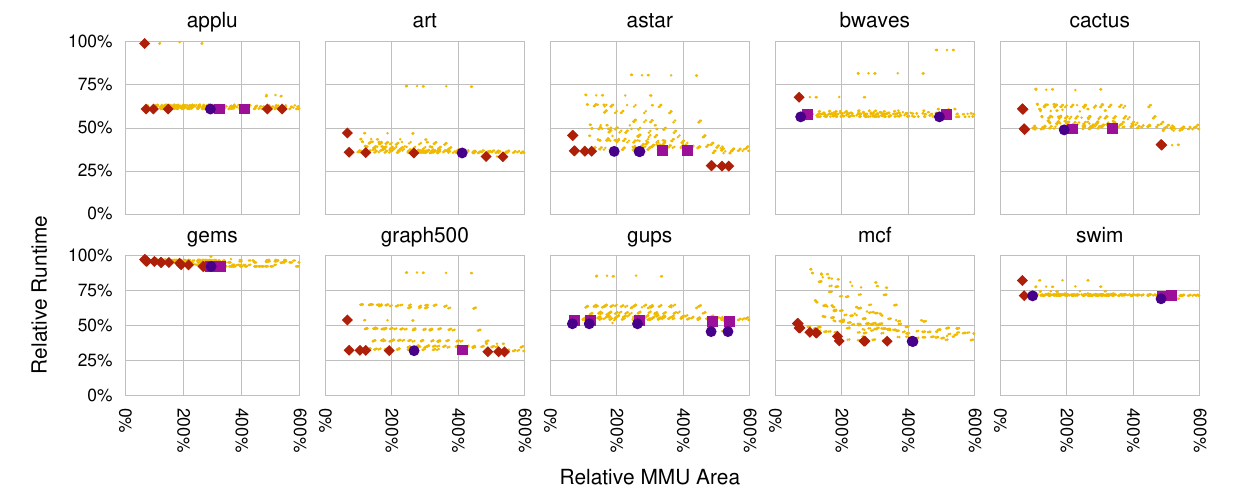}
  \caption{The area- and energy-optimal system-level MMU
    configurations rarely overlap, as small resources area-wise, such
    as L1 TLBs have an outsize impact on energy.}
  \label{fig:expt3}
  \end{minipage}
  \hspace{0.5em}
  \begin{minipage}[t]{1.75in}
 \includegraphics{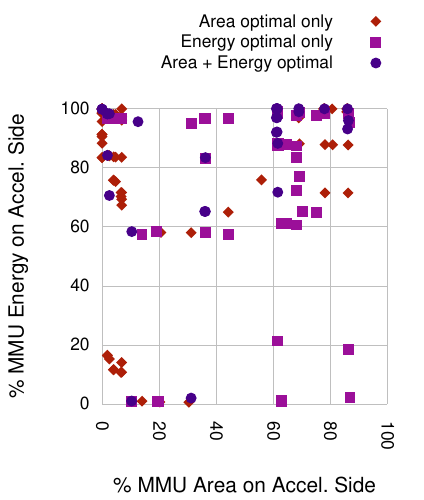}
  \caption{In optimal system-level MMUs, the area and energy splits between the CPU and accelerator are uncorrelated.}
  \label{fig:expt3-breakdown}
  \end{minipage}
\end{figure*}

\mypara{Mind the gap.}  The most surprising takeaway from this design
space exploration, as seen in \figref{expt2}, is the often-substantial
gap between the lower bound on accelerated runtime (Amdahl's bound)
and an ideal L1 TLB with 100\% hits.  This gap, as high as 40\% in
some cases, arises because even a perfect L1 TLB does have a hit time,
and even that can become a bottleneck for the
accelerator. Reclaiming this performance loss requires a fundamental
rethink of address translation. 

%% For example, it may inspire schemes that remove address translation
%% from the critical path of memory accesses and overlap them with other
%% work \cite{austin:isca96, xue:islped13}. We believe that these ideas
%% are largely unexplored and present a fruitful area of future research.

\mypara{MMU structural analysis.}  Analyzing the area-performance
tradeoffs of these 54 MMU configurations, we derive
several general insights.

In limited circumstances, TLB-less MMUs with only a page table walker
can be viable design options -- i.e., the address translation time
does not entirely negate the acceleration speedups.  However, not only
do PTWs have to be heavily multithreaded -- e.g., bwaves requires a
32-way PTW -- and thus fairly large, there are still many benchmarks
-- e.g., applu -- where in the best case, the PTW-only MMU remains
more than 20\% away from the ideal acceleration.  In fact, astar,
cactus, and gups with 32-way parallelized PTWs outperform the
idealized L1 TLB, with the parallel walks of well-cached page table
entries beating many small L1 TLB hits.  This suggests that PTW
organization is crucial to overall performance, with multithreading
being an important design variable.

Despite the cases described above, the addition of an L1 TLB typically
improves performance beyond a bare PTW.  For example, applu, art, and
swim make such effective use of an L1 TLB that they achieve close to
the ideal TLB ideal, even when paired with small serial PTWs.  In
other benchmarks, such as mcf, graph500, and cactus, the benefits are
more gradual, with roughly a 50\% performance improvement from the
smallest to the largest L1 TLBs.  Here, PTW multithreading remains
important, particularly when the L1 TLB is small.  Overall, these
configurations indicate that L1 TLBs can often boost performance, but
the gains quickly become incremental as the L1 TLB grows.

Finally, we see that L2 TLBs provide improved performance for little
area.  For example, the insertion of even a 256-entry L2 TLB into any
hierarchy gives an automatic 5-10\% performance boost.  However, as
with the L1, the benefits of growing the L2 are incremental, with a
1024-entry L2 TLB outperforming a 512-entry TLB by just
2-3\%. Overall, for a given area budget, it appears more
performance-effective to invest in PTW multithreading rather than
growing beyond 512-entry L2 TLBs.

\mypara{Impact of reference patterns.}  Finally, we assess the impact
of memory reference pattern reordering on these
results. \figref{expt2-patts} picks four benchmarks (astar, cactus,
gups, and mcf) which cover the range of observed behavior when access
patterns change. In general, the streaming reference pattern offers
the most benefit when total accelerator MMU area is small. This makes
sense since larger MMUs attain higher reach and process page table
walks more efficiently, making them less sensitive to reference
pattern. At larger MMU sizes, the original program order and streaming
orders tend to converge. However, random access remains noticeably
worse for astar and mcf, with atleast a 10\% performance gap in most
cases. Overall, however, these experiments suggest that the lower the
locality, the more valuable larger MMUs are.

%% file: expt3.tex
\subsection{System-level MMU Organization}
\label{sec:expt3}

In the final set of experiments, we explore area- and energy-efficient
system-wide MMUs -- i.e., MMUs that support both the CPU and an
accelerator.  

The results indicate that area and energy analysis of system-wide MMU
resources is nuanced and must take into account not only MMU resources
but interactions with other on-chip structures such as accelerators
and caches.  As in \secref{expt2}, we find that system-level MMU
design hinges on the workloads, suggesting there are careful strategic
decisions to be made as to which resources are private -- e.g., L1
TLBs -- and which are shared -- e.g., a large, heavily multithreaded
page table walker.

Furthermore, our analysis on the energy tradeoffs between additional
address translation hardware versus the cache accesses they eliminate
matches well-known observations that a large fraction of system energy
on today's chips is expended on data movement
\cite{keckler:micro11}. We find that in cases like address
translation, much of this data movement -- i.e., page table walks --
can be mitigated with TLBs.

Finally, although one might initially expect that MMU design choices
would change depending on the speedup enjoyed by the accelerator, we
found that the overall trends are in fact largely unchanged.
Streaming references, as expected, can get away with smaller MMUs,
expending lower energy, to achieve the same performance while
randomized references are slightly worse on both metrics.

%% In a similar spirit, it may be interesting to design MMUs that
%% match accelerator characteristics. For example, fixed function
%% accelerators may benefit greatly from private fixed-function MMUs,
%% whereas programmable units -- such as GPUs -- may employ
%% programmable MMUs like the ones that we have studied.

\mypara{System-level design space.}  As in \secref{expt2} we assume an
acceleration speedup of 10x and focus on the medium sized hotspots.
\figref{expt3} plots the area-performance trade-offs of 486
system-wide MMU configurations: the 54 accelerator MMUs explored in
the previous section crossed with 9 different CPU MMU configurations.
In this plot, we have highlighted the area-optimal designs, the
energy-optimal designs, and the designs that are both energy- and
area-optimal.

\mypara{MMU structural analysis.}  As is \figref{expt3} shows, the
energy- and area- Pareto optimal designs only rarely overlap.  It
turns out that the area and energy impacts of MMU components are very
different.  For example, adding even a small TLB for the accelerator
dramatically reduces the number of page table walks, which
significantly reduces the page table walk's L1, L2, and last-level
cache accesses.  Since caches -- particularly large multi-megabyte
ones -- expend more access energy than smaller 32-entry or 64-entry
TLBs, the overall area increases with the addition of TLBs, but the
overall energy decreases.  Only when TLBs are so overprovisioned that
they no longer conserve energy, do energy and area rise together.

\mypara{CPU and accelerator resource split.}  \figref{expt3-breakdown}
reveals how total MMU area and energy for the optimal designs is split
between the CPU and accelerator sides.  We see little pattern in the
split.  Spending an increased share of system-wide MMU area on the
accelerator side does not correlate with an increase in address
translation energy on the accelerator side.  For example, gems, which
executes few of its instructions on the accelerator, devotes most of
its energy to the CPU MMU, until the accelerator MMU gets very large,
expending a lot of energy to improve performance just incrementally.
This also reinforces the earlier observation that area
and energy behave quite differently as resources.

%% file: recommend.tex
\section{General Strategies and \\ Recommendations}
\label{sec:recommend}

This design space exploration has yielded a number of insights into
the dynamics of heterogeneous MMUs.

\takeaway{Relying on the CPU MMU to support a unified address space across
  accelerators of any granularity, but particularly small, is not a
  viable option.}  The overhead of even infrequent translations with
  1ns accelerator-CPU communication will wipe out the benefits of
  using accelerator (\secref{expt1}).

\takeaway{For complete address translation on an accelerator, a page table
  walker alone, if sufficiently parallelized, can outperform
  hierarchies that include TLBs (\secref{expt2})}

\takeaway{Applications respond to MMUs on acclerators in very different
  ways, with each application prefering a different MMU configuration
  (\secref{expt2}).}  Moreover, even
  an ideal (100\% hit rate) L1 TLB can be a performance bottleneck on
  an accelerator.  To maximize performance, MMUs
  on accelerators are best designed as application-specific MMUs, as
  specialized to the application as the accelerator itself.

\takeaway{To achieve good performance with a system-wide MMU area
  budget, accelerator and CPU MMU configurations must be carefully
  chosen with an understanding of the interaction among various
  components -- e.g. TLBs and page table walkers -- across the CPU and
  accelerators (\secref{expt3}).} Among area-optimal designs, accelerator
  MMUs consume a significant portion of the area.

\takeaway{While one may expect larger MMUs to expend more energy, this is
  usually not the case because large TLBs eliminate energy-intensive
  memory references from page table walks
  (\secref{expt3}).} Therefore, energy-optimal MMU
  configurations are often different from area-optimal designs.